\documentclass[11pt]{article}
\usepackage{amsmath,amssymb,color}

\usepackage{amsfonts}

\newcommand{\hoch}[1]{$\, ^{#1}$}

\newcommand{\be}{\begin{equation}}
\newcommand{\ee}{\end{equation}}
\newcommand{\bea}{\setlength\arraycolsep{2pt} \begin{eqnarray}}
\newcommand{\eea}{\end{eqnarray}}
\newcommand{\nn}{\nonumber}

\def\ft#1#2{{\textstyle{\frac{\scriptstyle #1}{\scriptstyle #2} } }}
\def\fft#1#2{{\frac{#1}{#2}}}

\def\0{{\sst{(0)}}}
\def\1{{\sst{(1)}}}
\def\2{{\sst{(2)}}}
\def\3{{\sst{(3)}}}
\def\4{{\sst{(4)}}}
\def\5{{\sst{(5)}}}
\def\6{{\sst{(6)}}}
\def\7{{\sst{(7)}}}
\def\8{{\sst{(8)}}}
\def\sst#1{{\scriptscriptstyle #1}}
\def\oneone{\rlap 1\mkern4mu{\rm l}}
\def\ep{{\epsilon}}
\def\del{{\partial}}

\begin{document}

\begin{flushright}
\hfill{MIFPA-11-19\ \ \ KIAS-P11028 }
\end{flushright}

\vspace{5pt}
\begin{center}
{\large {\bf Pseudo-supersymmetry, Consistent Sphere Reduction and\\
Killing Spinors for the Bosonic String}}

\vspace{7pt}

H. L\"u\hoch{1,2}, C.N. Pope\hoch{3,4} and Zhao-Long Wang\hoch{5}

\vspace{7pt}

\hoch{1}{\it China Economics and Management Academy\\
Central University of Finance and Economics, Beijing 100081, China}

\vspace{7pt}

\hoch{2}{\it Institute for Advanced Study, Shenzhen
University\\ Nanhai Ave 3688, Shenzhen 518060, China}

\vspace{7pt}

\hoch{3}{\it George P. \& Cynthia Woods Mitchell  Institute
for Fundamental Physics and Astronomy,\\
Texas A\&M University, College Station, TX 77843, USA}

\vspace{7pt}

\hoch{4}{\it DAMTP, Centre for Mathematical Sciences,
 Cambridge University,\\  Wilberforce Road, Cambridge CB3 OWA, UK}

\vspace{7pt}

\hoch{5}{\it School of Physics, Korea Institute for Advanced Study,
Seoul 130-722, Korea}

\vspace{10pt}

\underline{ABSTRACT}
\end{center}

Certain supergravity theories admit a remarkable consistent
dimensional reduction in which the internal space is a sphere.
Examples include type IIB supergravity reduced on $S^5$, and
eleven-dimensional supergravity reduced on $S^4$ or $S^7$.
Consistency means that any solution of the dimensionally-reduced
theory lifts to give a solution in the higher dimension. Although
supersymmetry seems to play a role in the consistency of these
reductions, it cannot be the whole story since consistent sphere
reductions of non-supersymmetric theories are also known, such as
the reduction of the effective action of the bosonic string in any
dimension $D$ on either a 3-sphere or a $(D-3)$-sphere, retaining
the gauge bosons of $SO(4)$ or $SO(D-2)$ respectively.  We show that
although there is no supersymmetry, there is nevertheless a natural
Killing spinor equation for the $D$-dimensional bosonic string.  A
projection of the full integrability condition for these Killing
spinors gives rise to the bosonic equations of motion (just as
happens in the supergravity examples).  Thus it appears that by
extending the notion of supersymmetry to ``pseudo-supersymmetry'' in
this way, one may be able to obtain a broader understanding of a
relation between Killing spinors and consistent sphere reductions.

\vspace{15pt}

\thispagestyle{empty}





\newpage

Kaluza-Klein dimensional reduction was introduced in the 1920's in
an attempt to unify four-dimensional gravity and electromagnetism
into the theory of pure gravity in five dimensions. Its most
important applications in physics came after the discovery of string
theories, whose natural space-time dimensions are higher than four.
As in the original motivation of Kaluza and Klein, dimensional
reduction can provide a natural interpretation for lower-dimensional
gauge symmetries as general coordinate transformations in the higher
dimension. In particular, the gauge group of the lower-dimensional
theory is associated with the isometry group of the internal space.

An important question that arises in a Kaluza-Klein reduction is
whether the procedure is consistent or not.  By consistency, we mean
that all solutions of the lower-dimensional theory are also
solutions of the higher-dimensional theory.  (Consistency is always
guaranteed if one retains the full Kaluza-Klein towers for all
modes; the issue here, though, is whether there exists a useful
consistent truncation to a finite set of modes.)  In fact the
original proposal of Kaluza and Klein to unify gravity and
electromagnetism in five-dimensional pure gravity could be said to
be only partially successful, because the consistency of the
reduction requires that an additional massless scalar field (the
dilaton) must be retained in the reduced four-dimensional theory.
The consistency in this, and many cases, can be understood
straightforwardly by a group-theoretic argument.  If the internal
space is symmetric under some group action, then it is consistent to
perform a dimensional reduction that retains all the singlets, and
only the singlets, under this action \cite{dufpop}.  Such a
reduction was called a DeWitt reduction in \cite{cvgilupo}. A simple
example is an $n$-torus reduction; it is consistent to keep all the
massless modes, since they are singlets under the action of the
$U(1)^n$ isometry group. More complicated examples, introduced by
DeWitt \cite{dewitt}, involve dimensional reduction on a group
manifold $G$, in which only those modes that are invariant under the
left action of the $G\times G$ isometry group are retained.

   A much more subtle reduction is exemplified by Pauli's (albeit
unsuccessful) attempt in the early 1950's to obtain $SO(3)$
non-abelian gauge fields by reducing six-dimensional gravity on the
2-sphere (see \cite{pauli,cvgilupo}).  The inconsistency in this
case can be understood by considering the untruncated theory in four
dimensions, prior to setting any of the fields in the Kaluza-Klein
towers to zero.  In this reduction, the $SO(3)$ gauge fields act as
sources not only for gravity, but also for certain {\it massive}
spin-2 fields in four dimensions.  Thus the massive spin-2
Kaluza-Klein tower cannot be consistently truncated in this
reduction.  In fact this same problem, of the retained gauge fields
acting as sources for massive spin 2 fields that one wants to
discard, means that Pauli reductions will, generically, be
inconsistent.  Remarkably, however, there do exist certain theories
for which a Pauli sphere reduction {\it is} consistent.

   Specifically, it has been demonstrated that in certain
supergravities where the theory admits an AdS$\times$Sphere vacuum,
it {\it is} consistent to perform a Pauli reduction on the
$n$-sphere in which  all the gauge fields associated with the
$SO(n+1)$ isometry group are retained. A notable example is the
$S^7$ reduction of $D=11$ supergravity \cite{dewnic}. Various other
examples of consistent sphere reductions of M-theory and type IIB
supergravity were obtained in \cite{nvv1}-\cite{Lu:2002uw}.  The
fact that there is a supergravity underlying the higher-dimensional
theory in all these examples might suggest that the consistency of
the reduction could be intimately related to the supersymmetry of
the higher-dimensional theory. Indeed, the demonstration of the
consistency of the $S^7$ reduction of eleven-dimensional
supergravity in \cite{dewnic} made extensive use of the Killing
spinors that exist in the pure AdS$_4\times S^7$ background.

As we shall discuss below, there also exist purely bosonic theories
that are not contained within any supergravities and that also admit
non-trivial consistent Pauli sphere reductions.  It is of
considerable interest to see if there exists any universal way of
characterising bosonic theories that admit Pauli reductions, to
encompass both the supersymmetric and the non-supersymmetric
examples.

One feature common to all the supergravity examples is that if one
looks at the equations for Killing spinors in purely bosonic
backgrounds, then by taking certain canonical projections of the
integrability conditions for the Killing spinor equations, one can
essentially derive the bosonic equations of motion for the theory.
For example, the gravitino transformation rule in bosonic
backgrounds in eleven-dimensional supergravity is $\delta\psi_M=\hat
D_M\ep$, where
\begin{equation}
\hat D_M= D_M -\ft1{288}(\Gamma_M{}^{N_1\cdots N_4}\, F_{N_1\cdots
N_4} - 8 F_{M N_1\cdots  N_4}\, \Gamma^{N_1\cdots N_4})\,,
\end{equation}
Projecting the integrability condition $[\hat D_M, \hat D_N]\ep=0$
with $\Gamma^M$ gives an equation of the form $(R_{MN} +
\cdots)\Gamma^N\ep=0$, where the factor in brackets vanishes by
virtue of the bosonic equations of motion. (This is related to
the fact that in the supersymmetry variation of
the action, the terms coming from varying $\psi_M$ in the gravitino
terms must cancel against those coming from varying the
bosonic terms in the action.) Thus, one may say that a
characterisation of the bosonic equations of motion in the
supergravity theories that admit consistent Pauli reductions is that
these equations can be derived from an appropriate projection of the
integrability condition for Killing spinors.

Recently, it was shown that for a large class of theories admitting
AdS$\times$Sphere vacua, encompassing the supergravities mentioned
above but including also non-supersymmetric theories, a broader
notion of Killing spinors can be introduced \cite{pseudo}.  In such
non-supersymmetric theories, bosonic backgrounds for which the
Killing spinor equations admit solutions were referred to as
``pseudo-supersymmetric.''\footnote{A related notion, referred to as
``fake supersymmetry,'' was introduced for scalar-gravity theories
in \cite{fake}.} The simplest class of such theories, which admit
AdS$\times$Sphere vacua, is provided by Einstein gravity coupled to
an $n$-form field strength, with the Lagrangian
\begin{equation}
{\cal L} = \sqrt{-g} (R - \ft{1}{2\, n!} F_n^2)\,,\label{genlag}
\end{equation}
where $F_n=dA_{(n-1)}$.   A Killing spinor
equation has been introduced for this system \cite{pseudo}, given
by
\begin{equation}
D_M \hat \epsilon + \fft{\tilde\alpha}{(n-1)!} \Gamma^{M_1\cdots
M_{n-1}} F_{M M_1\cdots M_{n-1}}\hat \epsilon +
\fft{\tilde\beta}{n!} \Gamma_M{}^{M_1\cdots M_n} F_{M_1\cdots
M_n}\hat \epsilon=0\,,\label{ksdef1}
\end{equation}
where $D_M$ is the covariant derivative, defined by
\begin{equation}
D_M\hat \epsilon \equiv \partial_M \hat \epsilon+ \ft14
(\omega_{M})^A{}_B \Gamma_A{}^B\hat \epsilon\,.
\end{equation}
The constants $(\tilde \alpha, \tilde \beta)$ are
given by
\begin{equation}
\tilde\alpha={\rm i}^{[(n+1)/2]}\, \fft{ \sqrt{\Delta}}{4d}\,,\qquad
d\tilde\alpha + \tilde d\tilde\beta =0\,,\label{ksdef11}
\end{equation}
where $d=n-1$, $\tilde d=D-n-1$, and $\Delta=2 d\tilde d/(D-2)$.

Although it was shown that the AdS$\times$Sphere vacuum, and a class
of $p$-brane solutions, are ``pseudo-supersymmetric'' with respect
to this definition of a Killing spinor, the integrability conditions
in (\ref{ksdef1}) are not in general consistent, in the sense that
the equations of motion following from  (\ref{genlag}) are necessary
but not sufficient to ensure the vanishing of the projected
integrability condition. Rather, additional constraints must still
be imposed \cite{pseudo}. These additional constraints are absent in
certain special cases, such as $n=4$, $D=11$, if a suitable $F\wedge
F\wedge A$ term is added to the Lagrangian; or in the case $n=5$,
$D=10$, if the 5-form is restricted to be self-dual. Interestingly
enough, these additional terms or restrictions are also precisely
what is needed in order to permit a consistent Pauli sphere
reduction.

Let us consider the case of ten-dimensional gravity coupled to a
5-form field strength in more detail. It was shown in \cite{pseudo}
that if one does not require the 5-form $H_\5$ in $D=10$ to be
self-dual, then the projected integrability condition for the
Killing spinor will only vanish upon use of the equations of motion
if, in addition, the extra constraints
\begin{equation}
H_{M_1[M_2M_3M_4M_5} H^{M_1}{}_{N_2N_3N_4N_5]}=0\,,\qquad
H_{M_1[M_2M_3M_4M_5} H_{N_1N_2N_3N_4N_5]}=0\label{d10n5int}
\end{equation}
are imposed. In \cite{pseudo}, a new class of pseudo-supersymmetric
``bubbling AdS geometries'' was constructed, that satisfy the
constraints (\ref{d10n5int}). In particular, as in the case of the
LLM solution \cite{LLM}, the new solution corresponding to the
elliptic disc boundary condition is expected to admit a reduction to
$D=5$, with an $S^5$ internal space. It is therefore of interest to
examine whether the extra conditions (\ref{d10n5int}) are related to
the consistency of the 5-sphere reduction.

The consistent $S^5$ reduction for $D=10$ with a self-dual 5-form
was obtained in \cite{clpst}.  The reduction ansatz is given by
\begin{eqnarray}
d\hat s_{10}^2 &=& \Delta^{1/2}\, ds_5^2 + g^{-2}\,
\Delta^{-1/2}\, T^{-1}_{ij}\,
  D\mu^i\, D\mu^j\,,\label{metans}\\
\hat H_\5 &=& \hat G_\5 + *\hat G_\5\,,\label{hans}
\end{eqnarray}
where
\begin{eqnarray}
\hat G_\5 &=& -g\, U\, \ep_\5 + g^{-1}\, (T^{-1}_{ij}\, {*D}
\, T_{jk})\wedge(\mu^k\, D\mu^i)\nn\\
&& -\ft12 g^{-2}\,
T^{-1}_{ik}\, T^{-1}_{j\ell}\, {*F_\2}^{ij}\wedge
D\mu^k\wedge D\mu^\ell\,,\label{gans}\\
{{\hat *}\hat G_\5} &=& \fft1{5!}\, \ep_{i_1\cdots i_6}\, \Big[
g^{-4}\, U\, \Delta^{-2}\, D\mu^{i_1}\wedge \cdots \wedge D\mu^{i_5}\,
\mu^{i_6}\nn\\
&& -5 g^{-4}\, \Delta^{-2}\, D\mu^{i_1}\wedge \cdots \wedge D\mu^{i_4}
\wedge DT_{i_5 j}\, T_{i_6 k}\, \mu^j\, \mu^k \nn\\
&&- 10 g^{-3}\, \Delta^{-1}\,
 F_\2^{i_1 i_2}\wedge D\mu^{i_3}\wedge D\mu^{i_4}\wedge D\mu^{i_5}\,
T_{i_6 j}\, \mu^j \Big]\,,\label{gdualans}
\end{eqnarray}
and
\begin{eqnarray}
&&U \equiv 2 T_{ij}\, T_{jk}\, \mu^i\, \mu^k -\Delta\, T_{ii}\,, \qquad
\Delta \equiv T_{ij}\, \mu^i\, \mu^j\,,\nn\\
&&F_\2^{ij} = dA_\1^{ij} + g\, A_\1^{ik}\wedge A_\1^{kj}\,,
\qquad DT_{ij} \equiv dT_{ij} + g\, A_\1^{ik}\, T_{kj} + g\, A_\1^{jk}\,
T_{ik}\,,\nn\\
&& \mu^i\, \mu^i = 1\,,\qquad
D\mu^i \equiv d\mu^i +g\,  A_\1^{ij}\, \mu^j\,,
\end{eqnarray}
with $\ep_\5$ being the volume form on the five-dimensional
spacetime. Note that ${{\hat *}\hat G_\5}$ is derivable from the
given expressions (\ref{metans}) and (\ref{gans}).  The coordinates
$\mu^i$, subject to the constraint $\mu^i\, \mu^i=1$, parameterise
points in the internal 5-sphere. It was shown in \cite{clpst} that
the reduction is consistent,  giving rise to lower-dimensional
equations of motion that can be derived from the five-dimensional
Lagrangian
\begin{eqnarray}
{\cal L}_5 &=& R\, {*\oneone} - \ft14 T^{-1}_{ij}\, {*D T_{jk}}\wedge
T^{-1}_{k\ell}\, DT_{\ell i} - \ft14 T^{-1}_{ik}\,
T^{-1}_{j\ell}\, {* F_\2^{ij}}\wedge F_\2^{k\ell}
-V\, {*\oneone}\label{d5lag}\\
&& \!\!\! - \ft1{48}\, \ep_{i_1\cdots i_6}\,
\Big(F_\2^{i_1 i_2}\, F_\2^{i_3 i_4}\, A_\1^{i_5 i_6} -
 g\, F_\2^{i_1 i_2}\, A_\1^{i_3 i_4}\, A_\1^{i_5 j}\, A_\1^{j i_6}
+\ft25 g^2\, A_\1^{i_1 i_2} \, A_\1^{i_3 j}\, A_\1^{j i_4}\,
A_\1^{i_5 k}\, A_\1^{k i_6} \Big)\,,\nn
\end{eqnarray}
where the potential $V$ is given by
\be
V = \ft12 g^2\, \Big(2 T_{ij}\, T_{ij} - (T_{ii})^2 \Big)\,.
\ee
(In (\ref{d5lag}), the wedge symbols in the final topological term
are omitted to economise on space.)

If instead we do not impose the self-duality condition for the
5-form, so that its reduction ansatz is now given simply by
\begin{equation}
H_\5 = \hat G_\5\,,\label{nonselfdual5form}
\end{equation}
then the reduction will not in general be consistent, since, as was
observed in
\cite{clpst}, the field equation $d{\hat *}G_\5=0$ gives rise to the
constraint
\begin{equation}
\ep_{ijk_1\cdots k_4}\, F_\2^{k_1 k_2}\wedge F_\2^{k_3 k_4}=0\,.
\label{ffterm}
\end{equation}
The intriguing point is that if we substitute the reduction ansatz
(\ref{nonselfdual5form}) into (\ref{d10n5int}), we arrive at exactly
the same constraint (\ref{ffterm}) that arose in \cite{pseudo} from
imposing the projected integrability condition for the Killing
spinors.\footnote{It is worth pointing out that there are large
classes of solutions in five dimensional gauged supergravity that
satisfy the condition (\ref{ffterm}). These solutions can now also
be lifted to the non-supersymmetric ten-dimensional theory where the
5-form is not self-dual.  A summary of such liftings, together with
an explicit example, is presented in appendix A.}

  Thus we find that the extra constraint needed for the projected
integrability of the Killing spinor for the $D=10$, $n=5$ system is
exactly the same as the extra constraint (\ref{ffterm}) that is
required for the consistency of the $S^5$ reduction.  This
observation leads us to speculate that the ability of a theory to be
consistently reduced, \`a la Pauli, on a sphere may go hand in hand
with its admitting some suitably-defined Killing spinor equation. In
some cases, namely certain supergravity theories, the Killing spinor
equation is simply the standard one associated with supersymmetry of
the bosonic background.  In more general situations, however, the
Killing spinor equation may be associated with a
``pseudo-supersymmetry'' that has not hitherto been considered.

There are some further examples that lend support to this idea.
Consider pure gravity in $(D+1)$ dimensions, for which the
Lagrangian is
\begin{equation}
{\cal L}_{D+1} = \sqrt{-\hat g} \hat R\,.\label{dp1lagpure}
\end{equation}
The associated Killing spinor equation is simply
\begin{equation}
0=D_M\hat \epsilon \equiv
\partial_M \hat \epsilon+ \ft14 (\omega_{M})^A{}_B \Gamma_A{}^B\hat
\epsilon\,,\label{puregravks}
\end{equation}
The projected integrability condition is
\begin{equation}
0=\Gamma^M [D_M,D_N] \hat \epsilon= \ft12 R_{MN}
\Gamma^M\hat\epsilon\,,
\end{equation}
which is satisfied by virtue of the Einstein equations of motion.
We now perform a Kaluza-Klein reduction on $S^1$, with the metric
ansatz given by
\begin{eqnarray}
d\hat s^2_{D+1} &=& e^{2\alpha\phi} ds^2_D + e^{2\beta\phi} (dz +
A_\1)^2\,,\cr 
\beta &=& - (D-2)\alpha\,,\qquad \alpha^2 =\ft{1}{2(D-1)(D-2)}\,.
\end{eqnarray}
The reduced $D$-dimensional Lagrangian in $D$ is
\begin{equation}
{\cal L}_D=\sqrt{g} (R - \ft12(\del\phi)^2 - \ft{1}{4} e^{a\phi}
F_\2^2)\,,\label{dlagf2}
\end{equation}
where $F_\2=dA_\1$ and $a=-2(D-1)\alpha$.  We can also perform the
Kaluza-Klein reduction of (\ref{puregravks}), to obtain the
equations for the $D$-dimensional Killing spinors:
\begin{eqnarray}
&&D_M\eta + \fft{\rm
i}{8(D-2)}e^{\fft12a\phi}\Big(\Gamma_M{}^{M_1M_2} -
2(D-2)\delta_{M}^{M_1}\Gamma^{M_2}\Big) F_{M_1M_2}\eta =0\,,\cr
&&\Gamma^M\partial_M\phi \eta + \ft{\rm i}{4}ae^{\fft12 a\phi}
\Gamma^{M_1M_2} F_{M_1M_2}=0\,.\label{Dks}
\end{eqnarray}
One can obviously expect that the projected integrability conditions
for these equations should be satisfied by virtue of the
$D$-dimensional equations of motion.  Indeed the projected
integrability conditions following from (\ref{Dks}) are given by
\begin{eqnarray}
&&\Big[R_{MN} - \ft12 \partial_M \phi \partial_N\phi - \ft12
e^{a\phi} (F_{MN}^2 - \ft{1}{2(D-2)} F^2 g_{MN})\Big]\Gamma^N \eta
\cr
&&-\ft{\rm i}{4(D-2)} e^{\fft12a\phi}\nabla_N F_{M_1M_2}\Big(
\Gamma_M\Gamma^{NM_1M_2} - 3(D-2) \delta_{M}^{[N} \Gamma^{M_1M_2]}
\Big)\eta \cr 
&&-\ft{\rm i}{2(D-2)} e^{-\fft12 a\phi} \nabla_N \Big(e^{a\phi}
F^N{}_{M_2}\Big) \Big(\Gamma_M \Gamma^{M_2} - (D-2)
\delta_M^{M_2}\Big)\eta=0\,,
\end{eqnarray}
and
\begin{eqnarray}
&&\Big(\nabla^2\phi- \ft1{4}ae^{a\phi}F^2\Big)\eta +\ft{\rm i}4
ae^{\fft12a\phi}\Gamma^{NM_1M_2} \nabla_N F_{M_1M_2}\eta\cr &&
+\ft{\rm i}{2} ae^{-\fft12a\phi} \Gamma^{M_2}
\nabla_N\left(e^{a\phi} F^N{}_{M_2}\right)\eta=0\,.
\end{eqnarray}
Thus the equations of motion imply that the projected integrability
conditions are satisfied for any dimension $D$.

The interesting point is that it is also consistent to perform a
Pauli $S^2$ reduction of the system (\ref{dlagf2}) in any dimension
$D$, yielding a theory in $(D-2)$ dimensions that includes the full
set of $SO(3)$ gauge bosons \cite{clpgen}.  This can be seen from
the fact that it is consistent to perform a (necessarily consistent)
DeWitt reduction of the pure gravity theory (\ref{dp1lagpure}) on
$S^3\sim SU(2)$, viewing it as the $SU(2)$ group manifold and
keeping all the singlets of the left-invariant action.  Since $S^3$
can be viewed as a $U(1)$ bundle over $S^2$, the reduction can be
split into two stages; an $S^1$ reduction followed by an $S^2$ Pauli
reduction.  Thus the consistency of the DeWitt reduction guarantees
the consistency of the Pauli $S^2$ reduction in this case
\cite{clpgen}.

    Of course, this is a rather simple example.  There are in fact
further examples of consistent Pauli sphere reductions of
non-supersymmetric theories. It was shown in \cite{clpgen} that it
is consistent to perform an $S^3$ or an $S^{D-3}$ Pauli reduction of
the effective action of the bosonic string in any dimension $D$.
This leads us to consider the possibility of a defining a Killing
spinor equation for the bosonic string.

The Lagrangian for the effective theory of the bosonic string in
$D$-dimensions is given by
\begin{equation}
{\cal L}_D= \sqrt{-g} \Big( R - \ft12(\partial\phi)^2 - \ft{1}{12}
e^{a\phi} F_\3^2\Big)\,,\label{stringlag}
\end{equation}
where $F_\3=dA_\2$ and $a^2=8/(D-2)$. The equations of motion are
given by
\begin{eqnarray}
\Box\phi &=& \ft1{12}a\, e^{a\phi} F_\3^2\,,\qquad
dF_\3=0=d(e^{a\phi} {*F_\3})\,,\cr 
R_{MN}&=& \ft12 \partial_M \phi \partial_N\phi + \ft14
e^{a\phi}\Big(F_{MN}^2 - \fft{2}{3(D-2)} F^2 g_{MN}\Big)\,.
\end{eqnarray}

We find that the appropriate equations for defining a Killing spinor
in this case are
\begin{eqnarray}
D_M\eta+ \ft{1}{96} e^{\fft12a\phi}\Big(a^2 \Gamma_M
\Gamma^{NPQ} - 12 \delta_{M}^N\Gamma^{PQ}\Big) F_{NPQ}\,\eta
&=&0\,,\label{ks1}\\
\Gamma^M\partial_M \phi\, \eta + \ft{1}{12}a e^{\fft12a\phi}
\Gamma^{MNP} F_{MNP}\,\eta&=&0\,. \label{ks2}
\end{eqnarray}
The forms of  these Killing spinor equations are motivated by
generalising the supersymmetry transformation rules for the
gravitino and dilatino in $D=10$, ${\cal N}=1$ supergravity
\cite{Bvn}.  The coefficients of each term are determined by
investigating the projected integrability conditions, whose
derivation is presented in appendix B.  They are given by
\begin{eqnarray}
&&\Big[R_{MN}-\ft{1}{2} \partial_M\phi\partial_N\phi
-\ft14e^{a\phi}(F^2_{MN}-\ft{2}{3(D-2)} F^2 g_{MN})\Big]\Gamma^{N}
\eta \cr 
&&-\ft{1}{6(D-2)} e^{\fft12a\phi}\nabla_N
F_{M_1M_2M_3}\left(\Gamma_{M}\Gamma^{NM_1M_2M_3}
-2(D-2)\delta_M^{[N}\Gamma^{M_1M_2M_3]}\right)\eta \cr 
&&-\ft{1}{2(D-2)}e^{-\fft12a\phi}\nabla_N \left( e^{a\phi}
F^N{}_{M_2M_3}\right)
\left(\Gamma_{M}\Gamma^{M_2M_3}-(D-2)\delta_M^{M_2}
\Gamma^{M_3}\right)\eta=0\,,\label{intcon1}
\end{eqnarray}
and
\begin{eqnarray}
&&\Big(\nabla^2\phi- \ft1{12}ae^{a\phi}F^2\Big)\eta
+\ft1{12}ae^{\fft12a\phi}\Gamma^{NM_1M_2M_3} \nabla_N
F_{M_1M_2M_3}\eta\cr && +\ft14ae^{-\fft12a\phi} \Gamma^{M_2M_3}
\nabla_N\left(e^{a\phi}
F^N{}_{M_2M_3}\right)\eta=0\,.\label{intcon2}
\end{eqnarray}
Thus we see that the projected integrability conditions are
satisfied by virtue of the full set of equations of motion. In the
special case when $D=10$, the theory can be supersymmetrised, to
give ${\cal N}=1$, $D=10$ supergravity, and the Killing spinor
defined above just reduces to the usual Killing spinor of the
supergravity theory. But the construction we have discussed here
works equally well in any spacetime dimension.

Sometimes it is advantageous to work with the theory in the string
frame, rather than the Einstein frame we have been using until now.
It is defined by rescaling the metric so that $ds^2_{\rm string} =
e^{-\fft12 a\phi} ds_{\rm Einstein}^2$. If we now define
$\Phi=-\phi/a$, the Lagrangian becomes
\begin{equation}
{\cal L} = e^{-2\Phi} (R + 4 (\partial\Phi)^2 - \ft{1}{12}
F_\3^2)\,.
\end{equation}
The defining equations for the Killing spinors, which are now scaled by the factor $e^{-\ft18a\phi}$, are given by
\begin{equation}
D_M(\omega_-) \eta=0\,,\qquad \Gamma^M \partial_M \Phi\,\eta -\ft{1}{12}
\Gamma^{MNP} F_{MNP} \eta=0\,,
\end{equation}
where $\omega_-$ is the torsionful spin connection, given by
\begin{equation}
\omega_{M\pm}{}^{AB} = \omega_{M}^{AB} \pm \ft12
F_{M}{}^{AB}\,.
\end{equation}

To conclude, we have observed an intriguing feature common to all
the known examples of consistent Pauli sphere reductions.  Namely,
in all such cases, the higher-dimensional theory admits a natural
definition of a Killing spinor.  A certain canonical projection of
the integrability conditions for the Killing spinor is satisfied by
virtue of the equations of motion of the theory.  In certain cases,
the projected integrability conditions may also impose quadratic
algebraic constraints on field strengths in the theory.  In such
cases, these turn out to be precisely the same as constraints that
must be imposed in order to achieve a consistent Pauli reduction.

We discussed various classes of examples that provide support for
this relation between consistent Pauli reductions and the existence
of a Killing spinor equation.  First of all, there are cases such as
eleven-dimensional supergravity and type IIB supergravity, where the
Killing spinor equation simply reduces to the standard Killing
spinor equations associated with supersymmetry.  We then considered
the example of ten-dimensional gravity coupled to a 5-form field
strength with no self-duality constraint.  In this case, we saw that
both the consistency of the Pauli $S^5$ reduction and the
consistency of the projected integrability conditions for the
Killing spinor equations required exactly the same quadratic
constraint (\ref{ffterm}) on the 5-form field.  Further examples
that we considered included dilatonic gravity coupled to a 2-form
field strength in any dimension, and dilatonic gravity coupled to a
3-form field strength in any dimension. The latter example arises as
the effective action for the bosonic string. The fact that there
exists a natural notion of a Killing spinor for the bosonic string
in an arbitrary spacetime dimension suggests that there may some
generalised geometric structure still to be uncovered.

\section*{Acknowledgement}

We are grateful to Haishan Liu and Paul Townsend for useful discussions.
The research of C.N.P. is supported in part by DOE grant DE-FG03-95ER40917.

\appendix

\section{Lifting of the solutions}

As discussed in the paper, the $S^5$ Pauli
reduction of the theory described by the Lagrangian
\begin{equation}
{\cal L}_{10} = \sqrt{-g} (R - \ft1{240} F_\5^2)\,,\label{n5d10lag}
\end{equation}
where $F_\5$ is not self-dual, is not in general consistent.  There
is an extra condition (\ref{ffterm}) that has to be satisfied.
However, this also implies that all the solutions of
five-dimensional theory (\ref{d5lag}) that satisfy (\ref{ffterm})
are also solutions of (\ref{n5d10lag}), with the lifting ansatz
given in this paper. Thus all the domain wall solutions supported by
the scalar fields, which are dual to the Coulomb branch of the
boundary conformal theory \cite{kltdomain,cglpdomain}, are solutions
of (\ref{n5d10lag}).  The $U(1)^3$ charged black holes in $D=5$
supergravity \cite{bcs} can be embedded not only in type IIB
supergravity \cite{10author}, but also in the theory described by
(\ref{n5d10lag}). Furthermore, the smooth $U(1)^3$ charged bubbling
soliton solutions obtained in \cite{clpbubble} can also be lifted
into solutions of (\ref{n5d10lag}). In particular, the single $U(1)$
charged solution can be lifted to give a pseudo-supersymmetric AdS
bubble geometry with an elliptic disc boundary condition, as was
constructed in \cite{pseudo}. Five-dimensional rotating black holes
do not in general satisfy the supplementary constraint
(\ref{ffterm}), and so they will not lift to give solutions of
(\ref{n5d10lag}). However, the singly-charged rotating black hole
constructed in \cite{zhiwei} does satisfy the condition
(\ref{ffterm}), and so in this case a lifting to give a solution of
(\ref{n5d10lag}) is possible. All such liftings use the reduction
ansatz we have given in this paper, and we shall not present them in
detail.

   Here, we present one simple
example in detail, namely the embedding of the five-dimensional
Reissner-Nordstr\o
black hole in (\ref{n5d10lag}).
Expressed in the notation we are using in this paper, the
five-dimensional Reissner-Nordstr\"om solution is given by
\begin{eqnarray}
ds_5^2 &=& -H^{-2}\, f\, dt^2 + H\, (f^{-1}\, dr^2 + r^2 d\Omega_3^2)
\,,\nn\\
T_{ij} &=& \delta_{ij}\,,\nn\\
A^{12} &=& A^{34}=A^{56}= \fft1{\sqrt3}\, A\,,
\end{eqnarray}
where
\begin{equation}
A= \fft{\sqrt{q(q+2m)}}{(r^2+q)}\, dt\,,\qquad
H= 1+\fft{q}{r^2}\,,\qquad f= 1-\fft{2m}{r^2} + g^2 r^2 \, H^3\,.
\end{equation}
Substituting into (\ref{metans}) and (\ref{gans}), we find that the
Reissner-Nordstr\"om solution lifts to give the ten-dimensional
solution
\begin{eqnarray}
d\hat s_{10}^2 &=& -H^{-2}\, f\, dt^2 + H\, (f^{-1}\, dr^2 + r^2
d\Omega_3^2)
+ g^{-2} (d\psi+ B-\fft{g}{\sqrt3}\, A)^2 + d\Sigma_2^2\,,\nn\\
\hat F_5 &=& 4g\, \ep_5 -\fft{1}{\sqrt 3\, g^2}\, {*F}\wedge J\,,
\end{eqnarray}
and hence
\begin{equation}
{\hat *}\hat F_5 = 2 g^{-4} \, (d\psi + B -\fft{g}{\sqrt3}\,
A)\wedge J\wedge J -\fft1{\sqrt3 g^3}\,  (d\psi + B
-\fft{g}{\sqrt3}\, A)\wedge F\wedge J\,,
\end{equation}
where $d\Sigma_2^2$ is the standard Fubini-Study metric on $CP^2$,
$J$ is its K\"ahler form, $dB=2J$, and $\psi$ is the coordinate on
the Hopf fibre over $CP^2$, with period $2\pi$.  (The proof of these
results follows using analogous manipulations to those in appendix B
of \cite{gilupapo1}.)

\section{Projected Integrability Conditions for the Bosonic String}

   Here, we derive the projected integrability conditions for the
Killing spinor equations for the $D$-dimensional bosonic string. We begin
by supposing that the Killing spinor equations take the form
\begin{eqnarray}
D_M\eta
+ a_2 e^{\fft12a\phi}\Big( \Gamma_M \Gamma^{M_1M_2M_3} - a_1
\delta_{M}^{M_1}\Gamma^{M_2M_3}\Big) F_{M_1M_2M_3}\,\eta
&=&0\, ,\label{pKS1}\\
\Gamma^M\partial_M \phi\, \eta - a_3 e^{\fft12a\phi}
\Gamma^{M_1M_2M_3} F_{M_1M_2M_3}\,\eta&=& 0\,\label{pKS2} .
\end{eqnarray}
The motivation for these equations is provided by the supersymmetry
transformation rules for the gravitino and dilatino in ten-dimensional
${\cal N}=1$ supergravity \cite{Bvn}.  The constants $a_1$, $a_2$ and
$a_3$ will be determined below.  We also leave the dilaton coupling
constant $a$ unspecified for now.

The next step is to compute the projected commutator $\Gamma^M
[D_{N},D_{M}]$ acting on $\eta$, and then to choose the undetermined
coefficients by requiring that it should vanish upon use of the
equations of motion.  After lengthy calculations, we find
\begin{eqnarray}
0&=&R_{MN}\Gamma^N\eta -2{a_2} e^{\fft12a\phi}\nabla_N
F_{M_1M_2M_3}\left(\Gamma_{M}{}^{NM_1M_2M_3}
-\ft43(a_1-3)\delta_M^{[N}\Gamma^{M_1M_2M_3]}\right)\eta\cr
&&-2a_2e^{-\fft12a\phi}\nabla_N \left( e^{a\phi}
F^N{}_{M_2M_3}\right) \left(3\Gamma_{M}{}^{M_2M_3}-2(a_1-3)
\delta_M^{M_2}\Gamma^{M_3}\right)\eta\cr
&&+\fft{aa_1a_2}{3a_3} \nabla_M\phi\nabla_N\phi\Gamma^{N}\eta
+2{a_2}\left(\ft23a_1-D+2\right)\nabla_{M}\left(
e^{\fft12a\phi}F_{M_1M_2M_3}\right)\Gamma^{M_1M_2M_3}\eta\cr
&&
+\Big[12(2a_1-3(D-4))a_2^2+9aa_2a_3\Big]e^{a\phi}
\Gamma_{M}{}^{M_2M_3N_2N_3}
F_{M_1M_2M_3}F^{M_1}{}_{N_2N_3}\eta \cr
&&+\Big[24(D-4)a_2^2-6aa_2a_3\Big]e^{a\phi}\Gamma_{M}F^2\eta \cr 
&&-\Big[4(4a_1-3(D-2))a_2^2+aa_1a_2a_3\Big]
e^{a\phi}\Gamma^{M_2M_3N_1N_2N_3} F_{MM_2M_3}F_{N_1N_2N_3}\eta\cr 
&&-\Big[8\left(2a_1^2-3(D-2)a_1+9(D-6)\right)a_2^2-6(6-a_1)aa_2a_3\Big]\cr
&&\qquad\qquad\times
e^{a\phi}\Gamma^{M_3N_2N_3}F_{M_1MM_3}F^{M_1}{}_{N_2N_3}\eta\cr
&&-\Big[8\left(2a_1^2-12a_1+9(D-2)\right)a_2^2-6aa_1a_2a_3\Big]
e^{a\phi}\Gamma^{N}F^2_{MN}\eta\,.
\end{eqnarray}
The vanishing of the $\nabla_{M}(e^{\fft12 a\phi}F_{M_1M_2M_3})$
term implies
\begin{eqnarray}
a_1=\ft32(D-2)\,,\label{a1sol}
\end{eqnarray}
which then leaves
\begin{eqnarray}
0&=&\Big[R_{MN}+\fft{aa_2(D-2)}{2a_3}
\nabla_M\phi\nabla_N\phi+\Big(24(D-4)a_2^2-6aa_2a_3\Big)e^{a\phi}
g_{MN}F^2\cr
&&-9(D-2)\Big(4(D-4)a_2^2-aa_2a_3\Big)e^{a\phi}F^2_{MN}\Big]
\Gamma^{N}\eta\cr 
&&-2{a_2} e^{\fft12a\phi}\nabla_N
F_{M_1M_2M_3}\left(\Gamma_{M}\Gamma^{NM_1M_2M_3}
-2(D-2)\delta_M^{[N}\Gamma^{M_1M_2M_3]}\right)\eta\cr 
&&-6a_2e^{-\fft12a\phi}\nabla_N \left( e^{a\phi}
F^N{}_{M_2M_3}\right)
\left(\Gamma_{M}\Gamma^{M_2M_3}-(D-2)\delta_M^{M_2}\Gamma^{M_3}
\right)\eta \cr 
&&+9(8a_2^2+aa_2a_3)e^{a\phi}\Gamma_{M}{}^{M_2M_3N_2N_3}
F_{M_1M_2M_3}F^{M_1}{}_{N_2N_3}\eta\cr 
&&-\ft32(D-2)(8a_2^2+aa_2a_3)e^{a\phi}\Gamma^{M_2M_3N_1N_2N_3}
F_{MM_2M_3}F_{N_1N_2N_3}\eta \cr 
&&-9(D-6)(8a_2^2+aa_2a_3)e^{a\phi}\Gamma^{M_3N_2N_3}F_{M_1MM_3}
F^{M_1}{}_{N_2N_3}\eta
\end{eqnarray}

   The terms involving $\nabla_M\phi\nabla_N\phi$,
$F^2_{MN}$ and $\Gamma_{M}{}^{M_2M_3N_2N_3}
F_{M_1M_2M_3}F^{M_1}{}_{N_2N_3}$ will then vanish upon use of the
equations of motion, provided that we choose
\begin{eqnarray}
&&a_3 + a\, a_2(D-2)=0\,,\qquad 8a_2^2+a\, a_2\, a_3=0\,,  \cr
&&9(D-2)(4(D-4)a_2^2-a\, a_2\, a_3)=\ft14\,,\label{arest}
\end{eqnarray}
for which the solution is
\begin{eqnarray}
a^2=\fft{8}{D-2}\,,\qquad a_2=\fft{1}{12(D-2)}\,,\qquad
a_3=-\ft1{12}a\,.\label{arest1}
\end{eqnarray}

   Acting on (\ref{pKS2}) with $\Gamma^N\nabla_N$, we have
\begin{eqnarray}
&&\nabla^2 \phi\, \eta =\Gamma^N\Gamma^M\nabla_N \nabla_M \phi\,
\eta =\Gamma^N D_N\left( \Gamma^M\partial_M \phi\,
\eta\right)-\Gamma^N \Gamma^M\partial_M \phi\,D_N\eta\cr
&=&a_3e^{\fft12a\phi}\Gamma^{NM_1M_2M_3} \nabla_N F_{M_1M_2M_3}\eta
+3a_3e^{-\fft12a\phi} \Gamma^{M_2M_3} \nabla_N\left(e^{a\phi}
F^N{}_{M_2M_3}\right)\eta \cr 
&&-6\left(D-2-\ft23a_1\right)a_2e^{\fft12a\phi}\Gamma^{M_2M_3}
F^N{}_{M_2M_3}\partial_N\phi\,\eta\cr 
&&+ \left[6(3D-2a_1-12)a_2a_3-\ft92a a_3^2\right]e^{a\phi}
\Gamma^{M_2M_3N_2N_3}F_{M_1M_2M_3}F^{M_1}{}_{N_2N_3}\eta\cr 
&&-  3\left[4(D-4)a_2a_3-a a_3^2\right]e^{a\phi}F^2\eta \,.
\label{pKS3}
\end{eqnarray}
This is also satisfied by the equations of motion, provided that the
coefficients $a$, $a_1$, $a_2$ and $a_3$ are chosen as in
(\ref{a1sol}) and (\ref{arest1}). Thus the Killing spinor equations
are given by (\ref{ks1}) and (\ref{ks2}).  The projected
integrability conditions are given by (\ref{intcon1}) and
(\ref{intcon2}).

Finally we would like to remark that we have investigated the
Killing spinors for a more general system with the 3-form field
strength replaced by an arbitrary $n$-form.  It turns out that that
projected integrability condition works only for two cases, the
bosonic string (\ref{stringlag}) and the Kaluza-Klein theory
(\ref{dlagf2}).

\end{document}